# HAMR Thermal Reliability via Inverse Electromagnetic Design

Samarth BHARGAVA and Eli YABLONOVITCH

University of California, Berkeley. USA

Abstract

Heat-Assisted Magnetic Recording (HAMR) has promise to allow for data writing in hard disks of beyond 1 Tb/in$^2$ areal density, by temporarily heating the area of a single datum to its Curie temperature while simultaneously applying a magnetic field from a conventional electromagnet. However, the metallic optical antenna or near-field transducer (NFT) used to apply the nano-scale heating to the media may self-heat by several hundreds of degrees. With the NFT reaching such extreme temperatures, demonstrations of HAMR technology observe write-head lifetimes that are orders of magnitude less than that required for commercial product. Hence, thermal reliability of the NFT is of upmost importance. In this paper, we first derive fundamental limits on the self-heating of the NFT to drive design choices for low temperature operation. Next, we employ Inverse Electromagnetic Design software, which provides deterministic gradient-based optimization of electromagnetic structures with thousands of degrees of freedom using the adjoint method. The Inverse Design software solves for unintuitive solutions to Maxwell's equations, and we present computationally generated structures for HAMR write-heads that offer a 40% or 170 °C reduction in NFT self-heating compared to typical industry designs.

I. Introduction

To ensure thermal stability of data over a 10 year lifetime in hard disks of beyond 1 Tb/in$^2$ areal density, the magneto-crystalline anisotropy of the magnetic granular media must be increased while scaling the magnetic grains to smaller dimensions [1]. However, recording information to such media is a monumental challenge. The current state-of-art employs writing electromagnets that are already limited by magnetic field saturation of permeable metals, placing an upper bound on the strength of magnetic field that can be applied during the recording process. Heat-Assisted Magnetic Recording (HAMR) has promise to allow for writing to highly anisotropic media, by temporarily heating the area of a single datum to its Curie temperature while simultaneously applying a magnetic field from a conventional electromagnet [2], [3]. In practice, a metallic optical antenna or near-field transducer (NFT) focuses light onto the highly absorbing magnetic recording layer in the media and locally heats a 30 x 30 nm$^2$ spot on the media to near 700 K [4], [5]. However, because the metal comprising the NFT, typically gold, is itself highly absorbing at optical frequencies, the NFT also heats by several hundreds of degrees [6]. This NFT self-heating is a significant cause of failure in HAMR systems and limits the lifetime of today's prototype HAMR write-heads to be orders of magnitude less than the desired 10 year lifespan [6]. Hence, an important Figure of Merit for reliability in a HAMR system is the temperature ratio between the media hotspot and NFT.

The difficulty in designing a thermally reliable HAMR system is two-fold: (1) the fundamental limits on the NFT temperature are not well known; (2) designing the optical system that produces nano-scale heating requires understanding the complex electromagnetic interactions of the illuminating waveguide, metallic NFT, magnetic write-pole and multi-layered magnetic media stack. On the first note, we derive in this paper a simple analytic model for the temperature rise in the NFT versus the temperature rise in the media hotspot. This provides important limits and constraints on the structural design of the NFT that must be satisfied for thermal reliability. On the second note, because of the wave nature of light, solutions to Maxwell's equations may be unintuitive and traditional design approaches based on intuition and highly constrained optimization through parameter sweeps can be inadequate. In this paper, we propose the use of Inverse Electromagnetic Design software, which provides deterministic gradient-based optimization of electromagnetic structures with thousands of degrees of freedom using the adjoint method [[7]–[9]]. The optimization of thousands of degrees of freedom are necessary because solutions to Maxwell's equations often have unintuitive and unconventional shapes that could not be designed by human intuition or analytic calculation alone. Gradient-based optimization is crucial for applications like HAMR, where a single 3D Maxwell simulation of nano-scale metallic structures is computationally demanding even on modern high-performance computing clusters. Because of the computational expense, heuristic algorithms like particle swarm and genetic algorithms that rely on randomness are too computationally burdensome for practical engineering design for optics in the nano-scale.

In this paper, our strategy towards achieving thermal reliability of the HAMR write-head was to prioritize design choices dictated by fundamental thermal limits and, then, to use our Inverse Electromagnetic Design software to find unintuitive solutions to provide optical performance in addition to the necessary thermal performance.

## II. Media - NFT Temperature Ratio

To understand the temperature ratio between the media and NFT, one may start with a simple model of spherical heat conduction from heat sources due to optical absorption in the media hotspot and the tip of the NFT, as shown in Fig. 1. In this simple model, we approximate the metallic NFT as a cone and the multi-layered media stack as a homogenous hemi-sphere. It is also assumed that there is not significant optical absorption and heat generation elsewhere in the media and NFT.

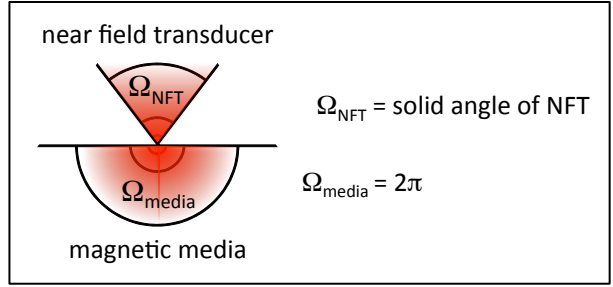

Fig. 1: Model of spherical heat conduction in a hemispherical media and conical NFT.

The temperature rise above ambient temperature in the NFT tip infinitely far away and equivalently the temperature rise in the media hotspot are described by

$$\Delta T_{NFT} = \frac{P_{NFT}}{\Omega_{NFT} \times K_{NFT} \times a_{min}} \qquad (1)$$

$$\Delta T_{media} = \frac{P_{media}}{\Omega_{media} \times K_{media} \times a_{min}} , \qquad (2)$$

where $P$ is the heat generated in the NFT tip or media hotspot, $\Omega$ is the solid angle, $K$ is the thermal conductivity of the materials and $a_{min}$ is the minimum diameter at the interface of the two structures. Then, we can relate the heat generated at the NFT tip and media hotspot to the optical absorption in these respective regions by

$$P_{NFT} = \omega \, \varepsilon_0 \, \varepsilon''_{NFT} \, |E_{NFT}|^2 \qquad (3)$$

$$P_{media} = \omega \, \varepsilon_0 \, \varepsilon''_{media} \, |E_{media}|^2 , \qquad (4)$$

where $\omega$ is the frequency of the excitation laser light, $\varepsilon_0$ is the free-space permittivity, $\epsilon''$ is the imaginary part of the permittivity and $|E|^2$ is the light intensity at the NFT tip or media hotspot. We can estimate the ratio of light intensity in the NFT tip to media hotspot according to electromagnetic boundary conditions at the NFT-air-media interface. Depending on whether the electric field is parallel or perpendicular to the interface, there are lower and upper bounds for this light intensity ratio given by equations 5 and 6 and depicted in Fig. 2. In this paper, we will assume the parallel boundary condition, as this is the worst-case scenario for a gold NFT and FePt recording layer at an excitation wavelength of 830 nm.

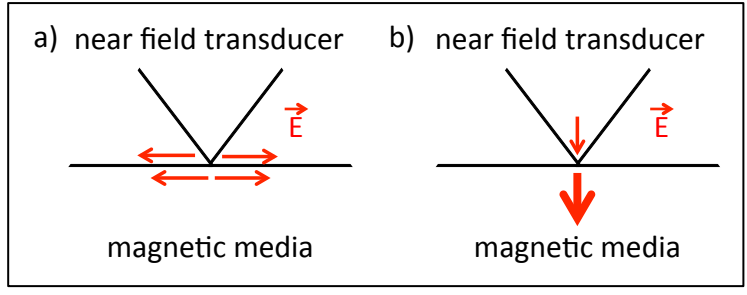

Fig. 2: Electromagnetic boundary conditions at the NFT-air-media interface for the two extreme cases, where (a) all the electric field is parallel to the interface and (b) all the electric field is perpendicular to the interface.

$$|E_{NFT}|^2 = |E_{media}|^2 \tag{5}$$

$$|\varepsilon_{NFT} E_{NFT}|^2 = |\varepsilon_{media} E_{media}|^2 \tag{6}$$

By combining the above equations, we can derive the following dimensionless ratio of the temperature rise in the media hotspot to the temperature rise of the NFT tip.

$$\frac{\Delta T_{media}}{\Delta T_{NFT}} = \frac{\varepsilon''_{media}}{\varepsilon''_{NFT}} \times \frac{K_{NFT}}{K_{media}} \times \frac{\Omega_{NFT}}{\Omega_{media}} \tag{7}$$

For a thermally stable NFT and write-head, this ratio must be as high as possible. Clearly, there are significant factors that are not accounted for in this expression, such as the anisotropic thermal conductivity of HAMR granular media and its under-layers, or the exact structural design of the NFT. However, this expression correctly emphasizes some key design choices for <u>low temperature NFT operation</u>.
1) The media should have the minimum amount of heatsinking.

2) NFT metallurgy should be optimized for $K_{NFT}/\varepsilon''_{NFT}$.

3) NFT structural design should include the largest solid angle of heat conduction at the tip of the NFT.

## III. Freeform Optimization via the Adjoint Method

Inverse Electromagnetic Design is based on two concepts: a) freeform optimization where the shapes are not constrained by a small number of parameters but rather thousands of parameters b) gradient-based optimization via the adjoint method to efficiently optimize the freeform shape. In context for HAMR, important electromagnetic Figures of Merit to be optimized may include the optical absorption integrated in the hotspot volume in the media, the ratio of absorption in the hotspot versus NFT tip, and the ratio of absorption in the hotspot versus secondary unwanted hotspots in the media. The gradient is the derivative of the chosen Figure of Merit with respect to all of the geometric parameters, which may be the continuous shape boundaries of the NFT and waveguide structures in the HAMR write-head. The gradient allows for the use of deterministic optimization algorithms like steepest decent, conjugate-gradient descent and Newton's method. In contrast, heuristic methods like genetic algorithms and particle-swarm optimizations rely on an element of randomness, whose computational burden is too cumbersome for applications in which even a single 3D simulation of Maxwell's equations are only feasibly on high-performance computing resources. The simplest but most inefficient method to calculating the gradient is finite-difference, which would require at least N+1 simulations, where N is the number of parameters. The adjoint method allows us to calculate the gradient with only 2 simulations regardless of the number of parameters and is a crucial part of our Inverse Design method. It will be described mathematically here in the context of electromagnetics.

First, let's denote the Figure of Merit (FOM) as the integral of an arbitrary function of electric field at locations $x$ within a particular volume $V_{FOM}$.

$$FOM = \int_{V_{FOM}} f\left(\vec{E}(x)\right) dx \tag{8}$$

The electric field in this region is a function of electromagnetic sources and of geometric structures, and we can model the electromagnetic effects of a small perturbation to the geometric structures. In this paper, we consider two possible structural perturbations. First, a *sparse* perturbation is the inclusion of an isolated small sphere of permittivity $\epsilon_2$ displacing a volume within a sea of permittivity $\epsilon_1$ as shown in Fig. 3a. Second, a *boundary* perturbation at the interface between two objects of permittivity $\epsilon_1$ and $\epsilon_2$ is the inclusion of a bump of $\epsilon_2$ replacing a volume of $\epsilon_1$ as shown in Fig. 3b. For either perturbation type, if the perturbation is electrically small, the electric field in this perturbed volume of $\epsilon_2$ is the same as the original electric fields in the displaced region of $\epsilon_1$, only differing by a different set of boundary conditions. For the *sparse* perturbation, applying boundary conditions around the perturbed sphere leads to equation 9, relating the electric field in the sphere to the original electric field in the sea of $\epsilon_1$ by the Clausius-Mossotti factor [10]. Similarly, for the *boundary* perturbation, we arrive at equation 10, which is the familiar boundary conditions at a flat interface, where $\parallel$ and $\perp$ denote the parallel and perpendicular vector components of the electric field. In both equation 9 and 10, $x'$ denotes the location of the perturbed volume.

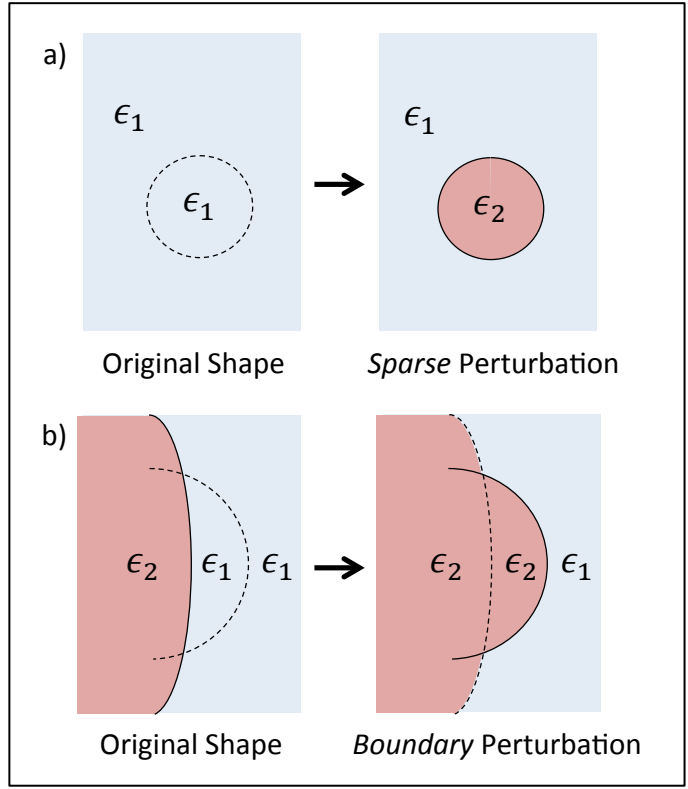

Fig. 3. We consider two possible perturbations to electromagnetic structures. (a) A *sparse* perturbation is the inclusion of an isolated small sphere displacing a material of different permittivity. (b) A *boundary* perturbation is the inclusion of a locally flat bump at the interface between materials of different permittivity.

$$\vec{E}_{perturbed}(x') \approx \frac{3}{\epsilon_2/\epsilon_1+2}\vec{E}_{orig}(x') \qquad (9)$$

$$\vec{E}_{perturbed}(x') \approx \vec{E}_{orig\parallel}(x') + \frac{\epsilon_1}{\epsilon_2}\vec{E}_{orig\perp}(x') \qquad (10)$$

The electromagnetic effects of these perturbations are effectively modeled by a change in dipole moment density in the perturbed volume as described in equations 11 and 12 for the *sparse* and *boundary* perturbations [7], [8], [11]–[13]. It is this additional electric source term that causes a change in electric field elsewhere in space. Of interest, the electric field in the volume where the FOM is evaluated is perturbed according to equation 13, where $G^{EP}(x, x')$ is electromagnetic Green's function relating a unit current source at the perturbation location $x'$ to the electric field induced at location $x$ within $V_{FOM}$. In application to the complex optical systems used for HAMR, this Green's function can only be evaluated by a full 3D Maxwell simulation with a current source at $x'$ and observing the numerically-calculated electric fields at $x$.

$$\frac{d\vec{P}_{pert}}{dV_{sparse}}(x') \approx (\epsilon_2 - \epsilon_1)\frac{3}{\epsilon_2/\epsilon_1 + 2}\vec{E}_{orig}(x') \tag{11}$$

$$\frac{d\vec{P}_{pert}}{dV_{bnd}}(x') \approx (\epsilon_2 - \epsilon_1)\left(\vec{E}_{orig\parallel}(x') + \frac{\epsilon_1}{\epsilon_2}\vec{E}_{orig\perp}(x')\right) \tag{12}$$

$$\vec{E}_{perturbed}(x) = \vec{E}_{orig}(x) + \vec{P}_{pert}(x')G^{EP}(x,x') \tag{13}$$

By differentiating equation 8 and using the chain rule, we arrive at an expression for the gradient, which is the derivative of the FOM with respect to a volumetric change in permittivity, shown in equation 14. The $2Re\{\ \}$ is a result of carefully taking the total derivative with respect to the complex valued functions $\vec{E}$ and $\vec{P}$, which is not shown in detail here for brevity. By differentiating equation 13, we can substitute the expression $[1(x')G^{EP}(x,x')]$ for $\frac{d\vec{E}(x)}{d\vec{P}_{pert}(x')}$, and we arrive at equation 15.

$$\frac{\partial FOM}{\partial V_{pert}(x')} = 2Re\left\{\int_{V_{FOM}}\frac{df}{d\vec{E}}(x) \cdot \frac{\partial \vec{E}(x)}{\partial \vec{P}_{pert}(x')} \cdot \frac{d\vec{P}_{pert}}{dV_{pert}}(x')\,dx\right\} \tag{14}$$

$$\frac{\partial FOM}{\partial V_{pert}(x')} = 2Re\left\{\int_{V_{FOM}}\frac{df}{d\vec{E}}(x) \cdot \left[\frac{d\vec{P}_{pert}}{dV_{pert}}(x') \cdot G^{EP}(x,x')\right] dx\right\}. \tag{15}$$

Note that when using equation 15, if we desire the unique value of the gradient at N possible boundary perturbations at locations $x'$, the term $\left[\frac{d\vec{P}_{pert}}{dV_{pert}}(x') \cdot G^{EP}(x,x')\right]$ must be evaluated by N individual Maxwell simulations of a current source equal to the dipole moment of each possible perturbation at each $x'$, respectively. Instead, we leverage reciprocity in electromagnetics, which Hendrik Lorentz proved in 1896 for any arbitrary arrangement of conducting or dielectric, isotropic or anisotropic bodies [14], which generically describes even the most complex optical structures used for HAMR. He proved that among such arbitrary structures, two current distributions $J_1$ and $J_2$ that separately induce the electric field distributions $E_1$ and $E_2$, respectively, are related by the following simple expression.

$$\int J_1 \cdot E_2\,dV = \int J_2 \cdot E_1\,dV \tag{16}$$

It is useful for us to consider the case of a unit current at $x'$ and the electric field it induces at $x$, in which case the well know reciprocity of electromagnetic Green's functions is obtained.

$$G^{EP}(x,x') = G^{EP}(x',x) \tag{17}$$

After substituting equation 17, we obtain a fundamentally different expression for calculating the gradient that is computationally inexpensive. In equation 18, the first term depends on the electric fields from a single *Forward* simulation of the original source and the unperturbed geometry. The latter term comprises the electric fields from a single *Adjoint* simulation where the source is the superposition of current sources of amplitude $\frac{df}{d\vec{E}}(x)$ in the volume of the FOM region. Hence, only 2 Maxwell simulations are required to obtain the gradient at all potential perturbation positions $x'$.

$$\frac{\partial FOM}{\partial V_{pert}(x')} = 2Re\left\{\frac{d\vec{P}_{pert}}{dV_{pert}}(x') \cdot \int_{V_{FOM}}\frac{df(x)}{d\vec{E}(x)}G^{EP}(x',x)\,dx\right\} \tag{18}$$

The above gradient calculation is generic and can be used for any type of geometric or electromagnetic perturbation of permittivity or electric field. By substituting equations 11 and 12, we arrive at the gradient calculations for the *sparse* and *boundary* perturbations that are used in this paper.

$$\frac{\partial FOM}{\partial V_{bnd}(x')} \approx 2Re\left\{\left[(\epsilon_2 - \epsilon_1)\left(\vec{E}_{orig\parallel}(x') + \frac{\epsilon_1}{\epsilon_2}\vec{E}_{orig\perp}(x')\right)\right] \cdot \left[\int_{V_{FOM}}\frac{df}{d\vec{E}}(x) \cdot G^{EP}(x',x)\,dx\right]\right\} \tag{19}$$

$$\frac{\partial FOM}{\partial V_{sparse}(x')} \approx 2Re\left\{\left[\frac{3(\epsilon_2 - \epsilon_1)}{\epsilon_2/\epsilon_1 + 2}\vec{E}_{orig}(x')\right] \cdot \left[\int_{V_{FOM}}\frac{df}{d\vec{E}}(x) \cdot G^{EP}(x',x)\,dx\right]\right\} \tag{20}$$

By using this gradient calculation, one can easily implement an iterative optimization using steepest descent,

where every iterative geometry update is in the direction of the gradient. In this work, we also used finite-difference between consecutive iterations to approximate the second derivative $\frac{\partial^2 FOM}{\partial V_{bnd}(x')^2}$, specifically the diagonal of the Hessian, to implement a quasi-Newton update method which demonstrated superior convergence as compared to steepest descent when the geometry contained thousands of parameters. Moreover, the freeform nature of the boundary optimization was implemented by representing the boundaries on a binary bitmap, where 1s and 0s represent the material inside and outside the various boundaries. Every pixel along the boundary was treated as a separate degree of freedom, and the boundary could expand outwards or contract inwards on a per-pixel basis. Hence, the optimized boundaries were allowed to completely diverge from the initial shape fed to the optimization algorithm. This freeform geometric representation combined with gradient-based optimization via the adjoint method allows for creative objective-first design of 3D electromagnetic structures, which we call *Inverse Electromagnetic Design*. The *Inverse Electromagnetic Design* software is available online for academic and commercial use at http://optoelectronics.eecs.berkeley.edu/PhotonicInverseDesign.

## IV. Maxwell Simulation Methods

The ability to perform accurate 3D electromagnetic models is imperative to computational optimization. Simulation results in this paper use a commercial finite-difference time-domain Maxwell solver, Lumerical FDTD, in which a pulse of light is injected into the waveguide of the HAMR system and propagated in the time domain towards the NFT, write pole and media stack structures until energy has decayed beyond our desired precision. A detailed mesh convergence test was performed to ensure minimal computational error due to discretization. The most crucial and computationally demanding mesh requirements were 1 nm cubic Yee cells in the metallic NFT and 0.5 nm Yee cell thicknesses in the media to resolve the various nanometer-thin layers in the media stack. We used an in-house high-performance computing cluster consisting 336 cores and 668 GBs RAM over 26 nodes. By parallelizing the solver through a Message-Passing Interface, Open MPI [15], and 40Gb/s Infiniband interconnects, our in-house cluster can simulate FDTD models of up to half a million Yee cell nodes. Typically, we ran simulations at the computationally optimal level of parallelization between 64 and 128 cores, with which we could run iterative optimizations of HAMR structures in roughly one day's time.

Fig. 5 shows 3D views of the HAMR structure that was modeled, and Table I contains structural and optical properties at the operation laser wavelength of 830 nm for the numerous write-head and media components, which was chosen to closely mimic designs from industry publications and patent literature. Table II shows thermal properties that were assumed for a thermal finite-element model performed in COMSOL Multiphysics to predict the media and NFT temperatures.

TABLE I
STRUCTURAL AND OPTICAL PROPERTIES IN PROPOSED HAMR SYSTEM

| Device | Dimensions | n | k |
|---|---|---|---|
| *Au NFT* | 125 nm radius, 60 nm thick | 0.16 | 5.08 |
| *Au NFT Peg* | 50 nm wide at ABS, 30 nm thick | 0.16 | 5.08 |
| *Ta$_2$O$_5$ Waveguide* | 100 nm thick | 2.1 | - |
| *SiO$_2$ Cladding* | - | 1.4 | - |
| *CoFe Writepole* | 120 wide at ABS | 3 | 4 |
| *Head Overcoat* | 2.5 nm thick | 1.6 | - |
| *Air Gap* | 2.5 nm thick | 1.0 | - |
| *Media Overcoat* | 2.5 nm thick | 1.2 | - |
| *FePt Recording Layer* | 10 nm thick | 2.9 | 1.5 |
| *MgO Interlayer* | 15 nm thick | 1.7 | - |
| *Au Media Heatsink* | 80 nm thick | 0.26 | 5.28 |
| *Media Substrate* | infinite | 1.5 | - |

TABLE II
THERMAL PROPERTIES IN PROPOSED HAMR SYSTEM

| Material | Specific Heat (J/m$^3$K) | Thermal Conductivity (W/mK) |
|---|---|---|
| Au | $3 \cdot 10^6$ | 100 |
| $Ta_2O_5$ | $2 \cdot 10^6$ | 2 |
| $SiO_2$ | $2 \cdot 10^6$ | 1 |
| CoFe | $3.5 \cdot 10^6$ | 20 |
| FePt - Lateral | $3 \cdot 10^6$ | 5 |
| FePt - Vertical | $3 \cdot 10^6$ | 50 |
| MgO | $2 \cdot 10^6$ | 3 |

### V. Inverse Electromagnetic Design Results

For low temperature operation, a crucial NFT design choice is to have the largest solid angle of thermal conduction from the tip of the NFT. The left of Fig. 4 shows a simplified HAMR system consisting of a magnetic write pole, media stack, gold lollipop NFT, and incident light through a waveguide mode similar to that of Seagate's parabolic solid-immersion mirror (PSIM) [6]. Through 3D FDTD modeling, we observe that this system produces a sharply confined hotspot in the media's recording layer, whose dimensions are defined by the cross-sectional dimensions of the NFT tip at the air-bearing surface (ABS). However, such an NFT has very little solid angle of heat conduction from the NFT tip, which may rise in temperature by hundreds of degrees and is a significant cause of failure in HAMR systems. Hence, we propose the structure shown on the right of Fig. 4 in which the entire top surface of the thin-film NFT structure directly contacts a bulk gold heatsink. The only part of the NFT that does not touch the bulk gold is the NFT tip because of the constraint that the magnetic write-pole must be adjacent to the NFT tip. As expected this new NFT structure behaves differently than the familiar lollipop NFT, and we observed that the incident mode that effectively excites the lollipop NFT, such as the PSIM, is not a good mode match to the proposed large solid angle NFT. The effect is observed by looking at the light intensity incident on the recording layer, shown in the bottom right of Fig. 4, in which the large solid angle NFT provides a poorly confined hotspot. This light intensity profile is unacceptable, because high temperatures in the media outside of the hotspot will unintentionally erase information. A typical data storage specification is to allow for 100,000 writes to a particular track without erasing data on nearby tracks. To meet this specification, the peak light intensity in the hotspot versus the peak intensity elsewhere in the media should be at least 5, and achieving this light intensity ratio with the proposed large solid angle NFT was the goal in this paper.

The strategy was to use UC Berkeley's Inverse Electromagnetic Design software to improve the mode match between the incident waveguide mode and the proposed NFT by optimizing an array of holes of low index material etched into the high index slab waveguide. A 3D perspective view of the proposed HAMR system is shown in Fig. 5. The incident light enters a slab $Ta_2O_5$ waveguide and evanescently couples to the NFT structure. The waveguide is patterned with holes of $SiO_2$ to reshape the incident mode to better couple to the proposed NFT. The gold NFT is sitting 15 nm above the waveguide and appears as a lollipop embossed directly underneath a bulk gold heatsink. A CoFe magnetic write-pole sits on top of the heatsink and the write-pole tip is 30 nm above the top surface of the NFT peg. Not shown in these 3D views is the magnetic media stack, described in Table I, which is adjacent to the right side of the waveguide, NFT peg and write-pole tip.

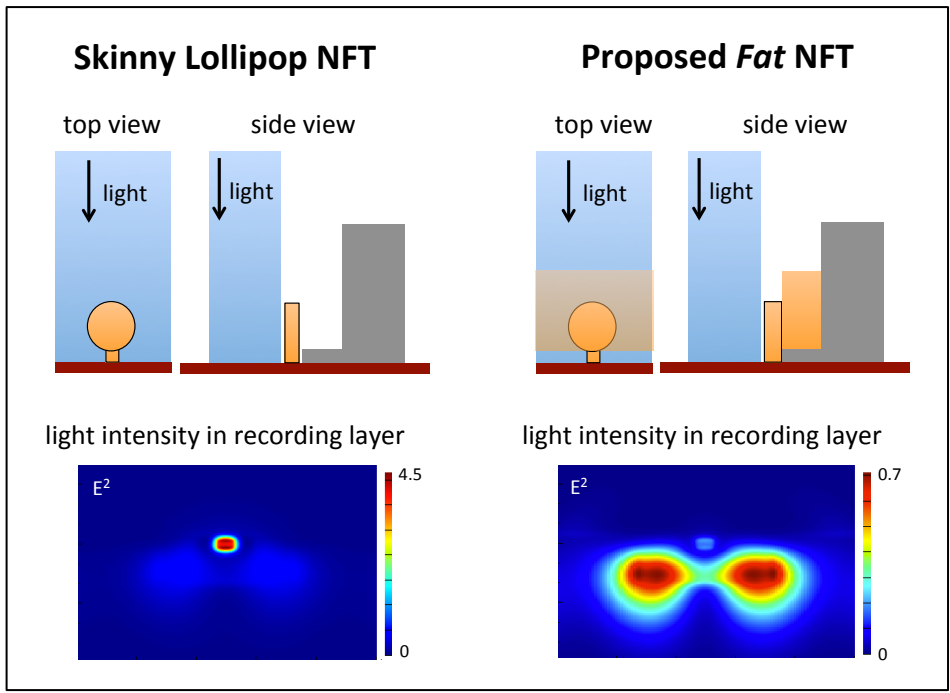

Fig. 4. A typical HAMR optical system composing of a $Ta_2O_5$ waveguide (blue), gold NFT (yellow), CoFe write-pole (grey) and magnetic media (red). On left, a typical lollipop NFT with no or little heatsinking produces a confined hotspot in the storage layer. On right, the proposed large solid-angle NFT without further design optimization is a poor mode-match to the same waveguide and offers little coupling to the hotspot.

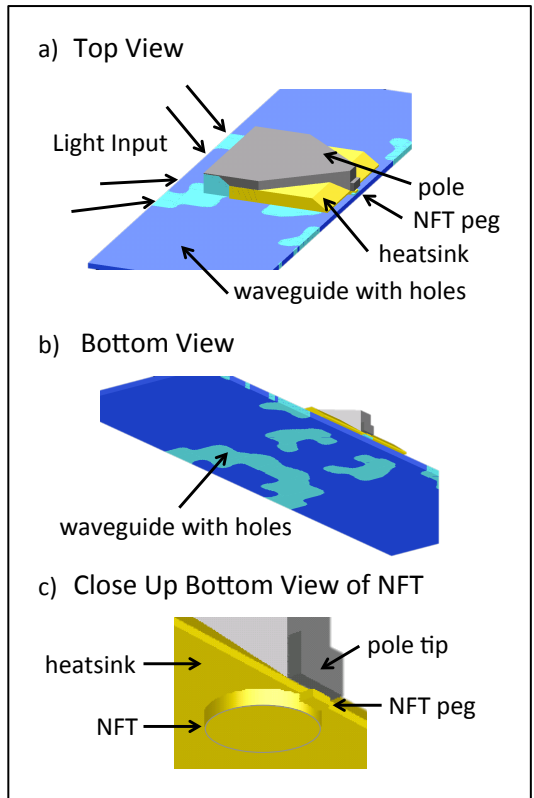

Fig. 5. 3D views of the proposed HAMR light delivery structure. The NFT is a lollipop embossed on a bulk chunk of gold, and the slab waveguide contains a pattern of low index material.

The FOM for the optimization was the light intensity ratio between the hotspot and unwanted sidelobes. The geometry that was optimized was a binary bitmap of 75,000 pixels, where each pixel represented an 3D voxel of either $SiO_2$ or $Ta_2O_5$ of dimensions 8 x 8 x 100 $nm^3$ occupying a total volume of 4 x 1.2 x 0.1 $\mu m^3$, which was the region of the slab waveguide under the NFT and adjacent to the ABS. Additional constraints on the binary bitmap were employed to enforce that the minimum diameter of a $SiO_2$ hole was greater than 128 nm and the radius of curvature of any boundary was at least 64 nm. Fig. 6 shows the iterative optimization of the holey waveguide pattern over 15 iterations, representing a total of only 30 simulations to optimize 75,000 degrees of freedom. In the first iteration, the software was configured to use the *sparse* gradient to take a huge step where it added many new $SiO_2$ holes into the $Ta_2O_5$ waveguide core. In the latter iterations, the software calculated the *boundary* gradient and was constrained to make boundary changes only. Fig. 7 shows a plot of the FOM versus iteration, showing smooth stable convergence towards a locally optimal design. Note, that between iterations 10 and 15, the geometry kept changing with little change to the FOM, suggesting that the optimal solution is robust to small variations in the boundaries of the waveguide pattern.

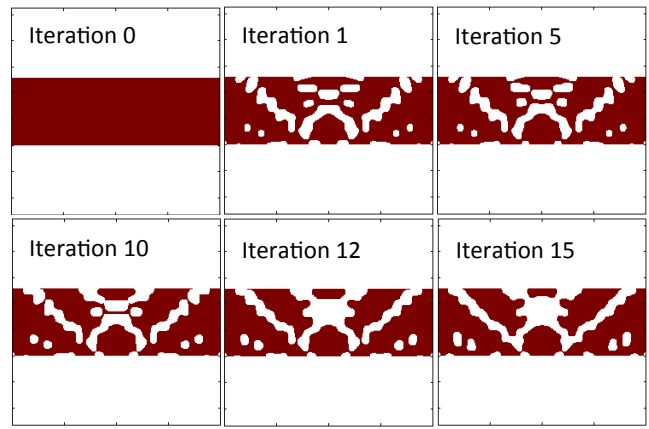

Fig. 6. Top view and iterative evolution of a $Ta_2O_5$ slab waveguide (red) patterned with $SiO_2$ holes (white). This unintuitive optimized pattern offered better optical coupling efficiency to the hotspot and reduced unintentional erasure of adjacent tracks.

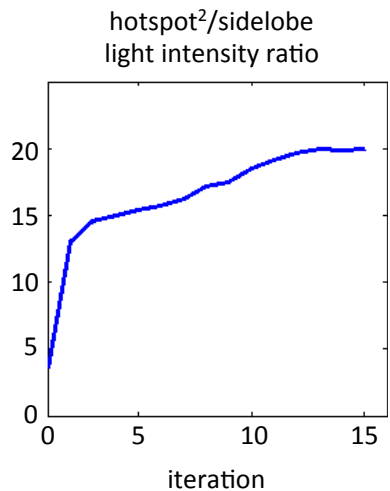

Fig. 7. Convergence plot of the optimized FOM versus iteration. The FOM was the square of the peak light intensity in the media hotspot divided by the peak light intensity in the unwanted sidelobes.

For a fair comparison, we modeled a typical heatsink structure for a lollipop NFT, shown in Fig. 8, that consists of a gold cylinder touching the center of the NFT and connected to the same bulk gold heatsink that is used in the proposed design. The narrowness of the cylindrical heatsink limits the heat conduction out of the NFT peg. We exported the optical absorption profile as a heat generation source in a thermal FEM model, in which we observed a peak temperature rise in the NFT peg of 440 °C above ambient when injecting enough light into waveguide to achieve a desired 400 °C temperature rise in the media hotspot. The optimized waveguide pattern coupled to the proposed large solid angle NFT is shown in Fig. 9. Using identical simulation models, we observed that the new proposed structure produces nearly identical optical properties of the media hotspot. Specifically, the hotspot shape is still well defined by the NFT peg dimensions, the power absorbed in the hotspot normalized to injected power into waveguide is ~6%, and the light intensity ratio between the hotspot and undesired sidelobes is greater than 5. Most importantly, from thermal modeling, we observed that the proposed NFT has a peak temperature rise in the NFT peg of only 270 °C above ambient when injecting enough light into waveguide to achieve the same 400 °C temperature rise in the media hotspot. This represents a 170 °C reduction or a ~40% lower temperature rise compared to a lollipop NFT with a typical cylindrical heatsink. This is expected, because the typical cylindrical heatsink is far away from the NFT peg and offers little solid angle of heat conduction. A 170 °C reduction in NFT temperature could result in massive improvements to HAMR write-head lifetimes.

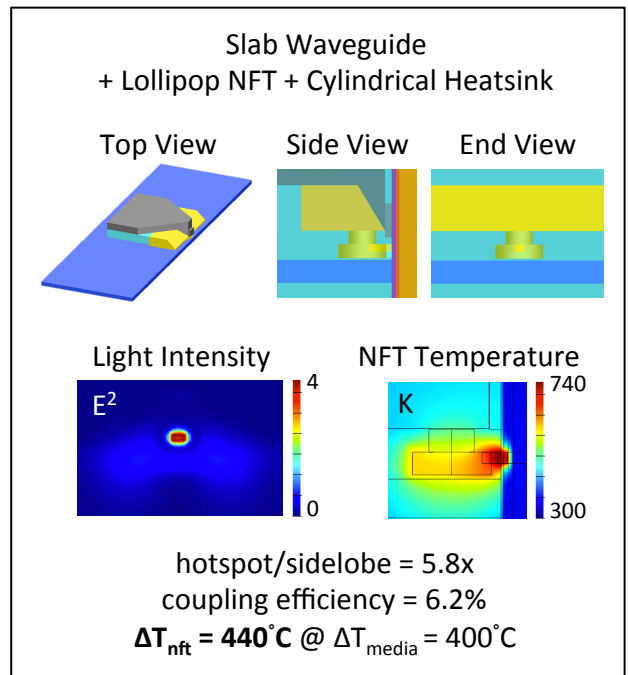

Fig. 8: (top) Structural model of an industry-like HAMR structure consisting of a slab waveguide, lollipop NFT, narrow cylindrical heatsink and write pole. (mid) Simulated light intensity in the recording layer and side-view temperature profile of the NFT, heatsink and media stack. This design achieves desirable optical properties but suffers from severe self-heating.

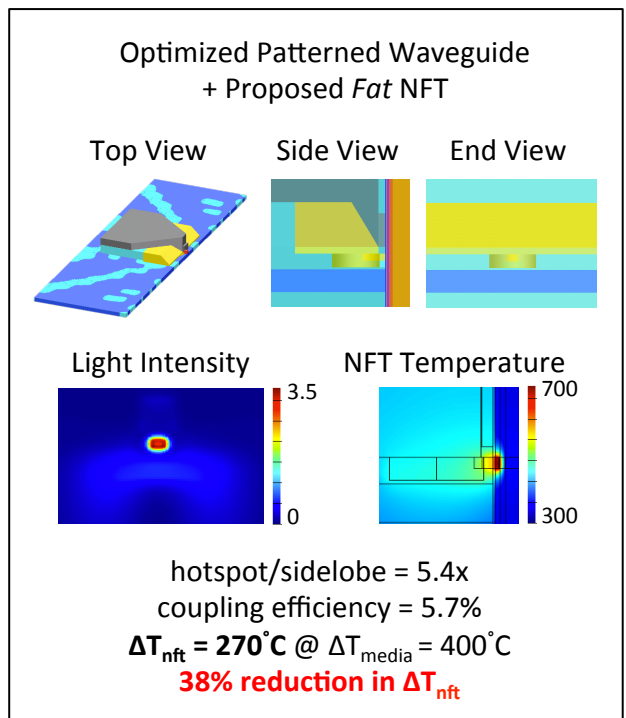

Fig. 9: (top) Structural model of the proposed HAMR structure consisting of a patterned waveguide, large solid angle NFT, and write pole. (mid) Simulated light intensity in the recording layer and side-view temperature profile of the NFT, heatsink and media stack. This design achieves desirable optical properties and significantly reduced self-heating.

## VI. Conclusions

The optical and thermal designs of a HAMR NFT are not problems that can be solved separately, because every object in the write-head affects both the electromagnetic and thermal physics. Prioritizing thermal design over optical design leads to unconventional NFTs like the large solid angle NFT proposed in this paper. With the power of Inverse Electromagnetic Design, we computationally generated unintuitive waveguide patterns that produced the desired optical properties in conjunction with the proposed *fat* NFT. Considering that reliability of structural and electronic devices often tends to vary with the exponential of temperature, the new structures proposed in this paper may offer orders of magnitude improvements to reliability by reducing the NFT self-heating by almost 40% or 170 °C compared to typical industry designs. Of course, the exact waveguide mode that illuminates the NFT and exact material properties of every device in the write-head and media system in commercial production must be accounted for in the optimization itself. Hence, the Inverse Electromagnetic Design software will be made available online at http://optoelectronics.eecs.berkeley.edu/PhotonicInverseDesign for the purpose that it may be directly applied to industry's custom HAMR designs. Future work will include simultaneous optimization of the NFT, heatsink and waveguide geometries. Also, co-optimization of numerous Figures of Merit may also provide significant advancements in computationally generated structures for reliable commercial HAMR technology.